\documentclass[twocolumn,aps,prb,showpacs]{revtex4}

\usepackage{graphicx}% Include figure files
\usepackage{dcolumn}% Align table columns on decimal point
\usepackage{bm}% bold math

\begin{document}

\preprint{APS/123-QED}

\title{H$_{\bm 2}$ in the interstitial channels of nanotube bundles}

\author{M.C. Gordillo}
\affiliation{Departamento de Ciencias Ambientales, Facultad de Ciencias
Experimentales, Universidad Pablo de Olavide, Carretera de Utrera km 1, 
41013 Sevilla, Spain}
\author{J. Boronat and J. Casulleras}%
\affiliation{Departament de F\'{\i}sica i Enginyeria Nuclear, Universitat
Polit\`ecnica de Catalunya, Campus Nord B4-B5, 08034 Barcelona, Spain 
}

\date{\today}

\begin{abstract}
The equation of state of H$_2$ adsorbed in the interstitial channels of a
carbon nanotube bundle has been calculated using the diffusion Monte Carlo
method. The possibility of a lattice dilation, induced by H$_2$ adsorption, has
been analyzed by modeling the cohesion energy of the bundle.
The influence of factors like
the interatomic potentials, the nanotube radius and the geometry of the
channel on the bundle swelling is systematically analyzed. 
The most critical input is proved to be
the C-H$_2$ potential. Using the same model than in planar graphite, which
is expected to be also accurate in nanotubes, the dilation is observed to
be smaller than in previous estimations or even inexistent. H$_2$ is highly 
unidimensional near the equilibrium density, the radial degree of
freedom appearing  progressively at higher densities.  
\end{abstract}


\maketitle

\section{INTRODUCTION}

Theoretical and experimental research in carbon nanotubes (CN's) is nowadays a
very active field in which new avenues are opening
continuously, all related to
its particular structure at the nanometer-scale level.
Besides their singular electrical and mechanical properties, CN's are able to
adsorb some atoms and molecules offering a real possibility for
quasi-one-dimensional environments. These are mainly the inner part of the
nanotubes and the interstitial channels (IC's) among them due to their natural
bundle arrangement. When CN's are formed, they appear as long capped cylinders
which only can adsorb a given substance if their caps are    
removed by chemical means. Instead, IC's are always present offering  very
narrow channels, distributed according to an hexagonal lattice, and with a
sizeable adsorption capability in comparison with adsorption in planar
graphite. Recently, Talapatra \textit{et al.} \cite{talapatra} studied 
experimentally the
binding energies of Xe, CH$_4$, and Ne on close-ended nanotube bundles and
concluded that none of them is adsorbed in the intersites. Probably, the
size of these atoms and molecules is too large compared with the radius of
the interchannel ($r_{\text{IC}} \simeq 3$ \AA) to fit into them. However, the
situation for lighter species like He or H$_2$ seems different. There are
several experiments which claim that both of them can be adsorbed in the
IC's
due to their small radius.\cite{teizer,williams,ye} H$_2$ is certainly the most 
interesting system
since a CN bundle  could be one of the best options to reach the
target energy densities for a lightweight hydrogen-storage system
usable for transportation. \cite{Dylon,dylon2}

In a recent theoretical work, Calbi \textit{et al.} \cite{Cole1} have studied
the adsorption of H$_2$, and other molecules and atoms, in the IC's of a CN
bundle. Introducing in the formalism a possible dilation of the bundle,
by means of a harmonic approximation, they concluded that in equilibrium the
bundle can swell. The dilation is observed in all cases,  with
different intensity depending on the particular system. In the geometry
there considered, corresponding to (10,10) tubes in the standard terminology,
and when the gas adsorbed is H$_2$ that swelling amounts approximately to 1
\%. This is the result of a delicate balance between two competing effects 
when the lattice is expanded: a decrease in the CN bundle cohesion energy and 
an increase in the H$_2$-nanotube binding energy. 
Obviously, this prediction for H$_2$ acquires special relevance since
it would imply a significant increase of the CN bundle storage power. 
However, at present there is not any experimental confirmation of this
possible adsorption-induced dilation. Probably, even if such effect is
indeed present, its manifestation could be hardly observed due to the
present experimental uncertainties. Raman spectroscopy seems a promising
method for a better insight. In a recent work, this method was applied
successfully by Williams \textit{et al.} \cite{williams} to discriminate the different 
adsorption sites of H$_2$ in CN bundles. 

In the present work, we present a diffusion Monte Carlo (DMC) study of
H$_2$ adsorbed in  IC's of a (10,10) CN bundle. In the past, and using the same
methodology, we characterized the ground-state properties of $^4$He and H$_2$ 
inside a single CN \cite{nano4,nano2} and of $^4$He adsorbed in an
IC.\cite{nanoic} In this paper, our aim is to
determine the equation of state of H$_2$ using microscopic theory, with
special attention to the possibility of an adsorption-induced bundle
dilation. To this end, the influence on swelling of inputs like the nanotube radius, the
molecule-molecule and CN-molecule interactions, the geometry of the
intersite, and the transversal degree of freedom are thoroughly discussed.
Our results agree somehow with the ones of Ref. \onlinecite{Cole1} but the
magnitude of a possible swelling is seen to be comparatively smaller or even
inexistent.

The rest of the paper is organized as follows. In Section II, the influence
of the several parameters of the model on a possible dilation of the bundle
is studied using a one-dimensional approximation. The accuracy of this
model is then tested in Section III by means of a three-dimensional DMC
calculation. Finally, Section IV comprises a brief summary and the main
conclusions.

\section{ONE-DIMENSIONAL APPROXIMATION: TESTING THE MODEL}

As in Ref. \onlinecite{Cole1}, we study H$_2$ adsorbed in the IC of a 
(10,10) CN bundle. The bundle is disposed in such a way that, in the plane 
perpendicular to the CN long axis, a triangular lattice is formed. The
lattice constant, i.e., the distance between
adjacent centers, is  $D_0 = 17$ \AA. $D_0$ corresponds to
the  equilibrium position and there is an overall agreement on its value.
Therefore, we consider it as a fixed parameter in all the calculations. 

Having established the value of $D_0$, the radius of a single nanotube
is directly related to the C-C minimum separation between two neighbors. 
In a theoretical calculation, Tersoff and Ruoff \cite{Tersoff}
concluded that this distance is  3.4 \AA , in agreement with  the most
accepted value for the Lennard-Jones $\sigma_{\text{C-C}}$. In fact, this
value coincides with the experimentally accepted distance between graphite sheets. 
Considering this C-C distance, the radius of a (10,10) tube is 6.8 \AA. 
It is worth mentioning that this  radius is the same one
obtains constructing the same nanotube by rolling up a graphitic surface
with a C-C distance of  1.42 \AA. However, there is not a general agreement
on those values. Recently, Charlier \textit{et al.} \cite{euro} carried out
a density functional calculation of a CN bundle and obtained an equilibrium
geometry corresponding to a C-C intertube distance of 
$\sim$ 3.2 \AA. The CN radius is then  6.9 \AA, a value which was used in
Ref. \onlinecite{Cole1}. At present and to our knowledge, there is not experimental data 
on the C-C
distance which could help to choose between different theoretical models. Therefore, we
consider in the present calculation two possible CN radius, 6.8 and 6.9 \AA.

The interstitial channel between three adjacent nanotubes in a (10,10)
bundle can lodge a hard sphere with radius  2.9 or 3 \AA, corresponding to
CN's with radius 6.8 and 6.9 \AA, respectively. Since the 
$\sigma$ parameter for the C-H$_2$ Lennard-Jones potential is around 3 \AA, 
one can reasonably assume a one-dimensional (1D) approximation for hydrogen 
adsorption in IC's. In this approximation, the H$_2$ energy per particle
($e=E/N$) can be written as \cite{Cole1}
\begin{equation}    
e(\lambda,D)= \epsilon_0 (D)+ e_{\text{1D}}(\lambda) + h(\lambda,D) + \frac{1}{\lambda}
\, \frac{3}{4} \, k (D - D_0)^2 \ , 
\label{ener1d}
\end{equation}  
with $\lambda$ the linear density of the H$_2$ molecules and $D$ the lattice parameter.
In Eq. (\ref{ener1d}), $\epsilon_0 (D)$ is the binding energy of a single
molecule in the IC, $e_{\text{1D}}(\lambda)$ is the 1D H$_2$ energy per
particle, $h(\lambda,D)$ corresponds to the interaction energy with H$_2$
in neighboring IC's, and the last term takes into account the cohesion energy
of the bundle when it is dilated from the equilibrium lattice distance
$D_0$. In the following, we analyze the several terms entering
Eq. (\ref{ener1d}) to disentangle which are the relevant inputs influencing
a possible dilation of the bundle.

Both $\epsilon_0 (D)$ and $e_{\text{1D}}(\lambda)$  are calculated using
the diffusion Monte Carlo (DMC) method.\cite{les,boro4he}  The DMC method
solves in a stochastic way the $N$-body Schr\"odinger equation in imaginary
time for the wave function $f({\bm R},t) = \psi({\bm R})\, \Psi({\bm
R},t)$,
\begin{eqnarray}
-\frac{\partial f({\bm R},t)}{\partial t} & = & -D {\bm \nabla}^2 
f({\bm R},t) + D {\bm \nabla} ({\bm F} \, f({\bm R},t)) \label{dmc1} \\
 & & + (E_{\text L}({\bm R}) -E) f({\bm R},t) \ , \nonumber
%\label{dmc1}
\end{eqnarray}
with $\psi({\bm R})$  a trial wave function introduced for importance
sampling. In Eq.  (\ref{dmc1}), $D = \hbar^2/2m$,
$E_{\text L}({\bm R}) = \psi({\bm R})^{-1} H \psi({\bm R})$ is the local energy,
and  ${\bm F}({\bm R}) = 2\, \psi({\bm R})^{-1}
{\bm \nabla} \psi({\bm R})$ is the so-called drift force
which guides the diffusion movement to regions where $\psi({\bm R})$ is
large.

The trial wave function $\psi$ used in the calculation of 1D H$_2$ is a
Jastrow one,
\begin{equation}
\psi^{\text{1D}}({\bm R}) = \prod_{i<j} f(r_{ij})  \ ,
\label{trial1d}
\end{equation}
with $f(r_{ij})= \exp \left( -0.5 \, \left( b/\,r_{ij} \right)^5
\right)$  a McMillan two-body correlation factor.
The parameter $b$ in $f(r)$ has been determined by means of
a variational Monte Carlo (VMC) optimization. 
Near the equilibrium density, the optimal
value is $b = 3.996$ \AA\ and it increases gradually with the density
$\lambda$ (at $\lambda = 0.358$ \AA$^{-1}$, $b =4.026$ \AA).

The trial wave function used in the calculation of $\epsilon_0 (D)$
includes two-body correlations with the three nanotubes surrounding the IC.
They are also of McMillan type,
\begin{equation}
\psi^{\text{IC}}({\bm R}) = 
\prod_{n=1}^3 \exp \left[ -\frac{1}{2} \left (\frac{a}{r_{n}} \right)^5
\right] \ ,
\label{trialic}
\end{equation}
$r_n$ being the distance of the hydrogen molecule to the 
center of any of the three tubes. 
%In rigor, this is not a many-body calculation that could have done
%with other methods, but it is nevertheless exact. 
The parameter $a$, optimized using VMC, varies from 
$a = 22.5$ \AA\ for zero dilation to $a = 21.5$ \AA\ for $D-D_0$ = 0.2
\AA. On the other hand, the optimal values for the parameter $a$ present a negligible
dependence with the particular C-H$_2$ potential chosen in the calculation.
The IC is so narrow that one can approximate the sum of individual
CN-molecule potentials by a new one which is only a function of the radial
distance $r$ to the center of the IC. This simplified model is obtained by an
azimuthal average of the three individual potentials.\cite{Cole1} In this case, the
trial wave function is simpler than the previous model (\ref{trialic}). We
have chosen a gaussian  
\begin{equation}
\psi_{\text{c}}({\bm R}) =  \exp(-c \, r_n^2) \ ,
\label{trialicr}
\end{equation}
the parameter $c$ varying from 7.6 \AA$^{-2}$ to 11.5 \AA$^{-2}$, depending
on the radius of the tube and on the dilation. The greater the radius,
the smaller the value of $c$.

Para-hydrogen is spherical to a large extent. As usual in most of molecular
hydrogen calculations, the intermolecular interaction is considered purely
radial and described by the Silvera and Goldman (hereafter SG) model.\cite{Silvera} In
addition, and to make comparisons with previous work (Ref.
\onlinecite{Cole1}),
we have also made some calculations with the potential proposed by Kostov,
Cole, Lewis, Diep, and Johnson.\cite{kostov}  This H$_2$-H$_2$ potential (hereafter KCLDJ) 
incorporates, in
a rather crude way, three-body corrections to the pair potential coming
from the triplets H$_2$-C-H$_2$.

Much more critical for estimating a possible bundle dilation is the model for
the CN-H$_2$ interaction. As in our previous work on the equation of state
of H$_2$ adsorbed inside a nanotube,\cite{nano2} we use the CN-H$_2$ potential proposed
by Stan and Cole.\cite{Stan} This potential results from an average
over all the C-H$_2$ interactions between the C of an infinite CN and a 
single molecule located at a radial distance $r$. 
The interaction is thereby independent of $z$ (corrugation effects are
neglected) and can be used for adsorption both inside and outside the nanotube. 
The resulting CN-H$_2$ potential depends explicitly on the $\sigma$ and 
$\epsilon$ parameters of the Lennard-Jones potential between C and H$_2$.
At present, there is not an overall agreement about which are the best set
of parameters ($\sigma$,$\epsilon$) describing this interaction. In order
to study the influence of this choice on the calculations, two different
sets have been studied. The first one, used for example in Refs.
\onlinecite{Cole1} and \onlinecite{uptake}, is derived with 
the Lorentz-Berthelot combining  rules
\begin{eqnarray}
\sigma_{\text{C-H}_2} & = &  \frac {\sigma_{\text{H$_2$-H$_2$}} +
\sigma_{\text{C-C} }}{2} \\
\nonumber
\epsilon_{\text{C-H}_2} & = & \sqrt {\epsilon_{\text{H$_2$-H$_2$}} \,
\epsilon_{\text{C-C} }} \ .
\label{rules}
\end{eqnarray}
Using $\sigma_{\text{C-C}} = 3.4$ \AA, $\epsilon_{\text{C-C}}=28$ K,
$\sigma_{\text{H$_2$-H$_2$}}
=3.05$ \AA, and $\epsilon_{\text{H$_2$-H$_2$}}=37.0$ K the first set is 
$\epsilon_{\text{C-H}_2} = 32.2$ K and $\sigma_{\text{C-H}_2} = 3.23$  \AA (hereafter LB). 
The second option is to consider the optimal parameters which describes 
the interaction of H$_2$ with planar graphite. This set from
Wang, Senbetu, and Woo \cite{wang} is probably more
realistic than the first one and it has also been used in the
past.\cite{williams,nano2,sc} The
values are  $\epsilon_{\text{C-H}_2} = 42.8$ K and $\sigma_{\text{C-H}_2} = 2.97$ \AA 
(hereafter WSW). 
Notice the sizeable difference between both sets which will generate
significant differences in the final results.

Dilation and compression of the lattice constant in the CN bundle have an
energetic cost. Following Refs. \onlinecite{Mizel} and \onlinecite{Cole1},
that contribution is assumed to be harmonic in the displacement around the
equilibrium position $D_0$ (see  Eq. \ref{ener1d}). This approximation is
expected to hold for small $D-D_0$ values but the uncertainty in the real 
value of $k$ is rather large. From compression modes measured in graphite
by Nicklow \textit{et al.},\cite{nicklow} Mizel \textit{et al.}
\cite{Mizel} estimated a value $k= 1740$ K \AA$^{-3}$. The constant $k$ can
also be obtained from the bulk modulus $B$. Considering that any deformation 
of the tube takes place only in a direction perpendicular to its long axis, 
$k =  2/\sqrt{3} \, B$.\cite{Tersoff} From the experimental measure of $B$ by  
Tang \textit{et al.} 
\cite{Tang} ($B=41.66$ GPa) one obtains $k= 3015 $ K \AA$^{-3}$, a figure much larger
than the one from Ref. \onlinecite{Mizel}. Part of the difference can
somehow emerge from the fact that the experimental data for $B$  was
obtained for CN's with wider radius (7.04 \AA). In the results presented
below, we have used $k= 1740$ K \AA$^{-3}$ everywhere but the influence on
the final results of the particular $k$ value is analyzed in selected
cases.
 
Finally, the term $h(\lambda,D)$ in Eq. (\ref{ener1d}) sums the interaction
energy between the H$_2$ molecules of a given IC and the ones of its
neighboring channels. This contribution may be readily obtained assuming  a
mean field approximation,
\begin{equation}
h(\lambda,D)  = \frac{\lambda}{2} \,  \int_{- \infty}^{\infty}  d x
\ V_{\text{H$_2$-H$_2$}} \left(\sqrt{x^2 + D^2}\right) \ ,
\label{mfield}
\end{equation}
which considers the neighboring channels as uniform arrays without
correlation effects that could modify the H$_2$ kinetic energy. 
The calculation in Eq. (\ref{mfield}) is extended to 
the nearest neighbors, the next-nearest
neighbors, and then on  up to the desired accuracy. Obviously,
$h(\lambda,D)$ changes when the bundle swells. In Ref. \onlinecite{yo}, we have
verified the high accuracy of Eq. (\ref{mfield}) by comparing its
estimation  with an exact DMC calculation.

\begin{table}
\centering
\begin{tabular}{cclc}
Radius (\AA) & IC & $V_{\text{C-H$_2$}}$ & $\epsilon_0$ (K) \\  \colrule

6.9 &     T    &     WSW   &   -1020.69(8)  \\
6.9 &     T    &     LB    &    -278.02(8)  \\

6.8 &     T    &     WSW   &   -1096.02(2) \\
6.8 &     T    &     LB    &    -668.39(9) \\

6.9 &     R    &     WSW   &    -965.93(7) \\ 
6.9 &     R    &     LB    &    -207.59(5) \\ 

6.8 &     R    &     WSW   &   -1044.49(2) \\ 
6.8 &     R    &     LB    &    -576.56(2) \\ 

\end{tabular}
\caption{Binding energy of  a single H$_2$ molecule adsorbed in 
the interstitial channels of a (10,10) CN bundle for different CN
radius, IC geometries, and C-H$_2$ potentials. Figures in parenthesis
are statistical errors.}
\end{table}

\begin{figure}
\centering
\includegraphics[width=9cm]{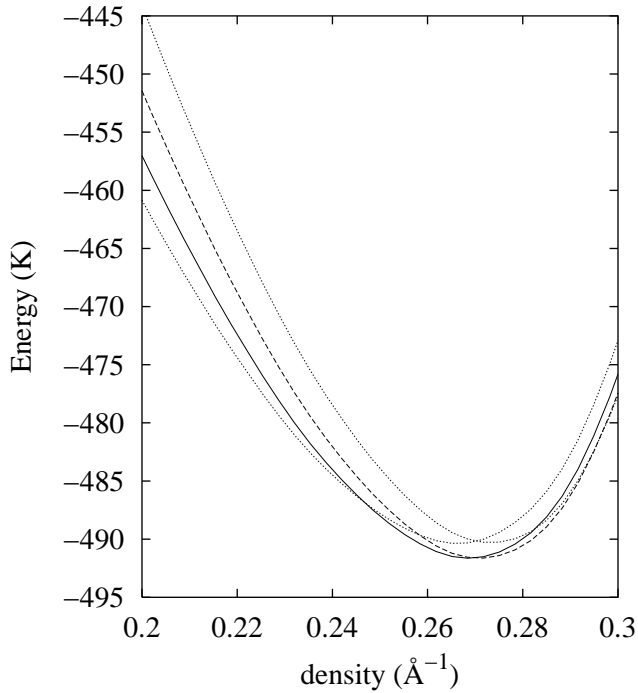}
\caption{\label{Fig 1} Energy per hydrogen molecule for CN radius 6.9 \AA,
T geometry, SG H$_2$-H$_2$  potential, and  LB C-H$_2$ potential.
>From top to bottom in $\lambda = 0.2$ \AA$^{-1}$: $D-D_0$ = 0.18,
0.17, 0.16 and 0.15 \AA.
}
\end{figure}

The influence of the CN radius and the C-H$_2$ potential in the equation of 
state of adsorbed H$_2$ is clearly observed in the value of $\epsilon_0$. 
Table I contains results for the H$_2$ binding energy considering two CN
radius, 6.8 and 6.9 \AA, the two C-H$_2$ potentials discussed above, LB
and WSW, and two geometries for the IC,
the real one with the triangular composition of the tubes surrounding the
IC (T) and the radial (R) model from an azhimutal average. The narrowness
of the IC makes the differences between the T and R results not to be
larger than 15 \%. The dominant effect is unquestionably the C-H$_2$
potential. The significant differences between the pairs
($\sigma$,$\epsilon$) for the two models generate a big discrepancy in the
value of $\epsilon_0$. The largest binding energies correspond to the most
accurate WSW potential due to its larger parameter $\epsilon$. It
is also noticeable, and will be discussed below in connection with
dilation, the
different behavior of $\epsilon_0$ with the CN radius: it strongly depends
on the radius for  LB  whereas remains nearly
unchanged for WSW. This last feature is a consequence of the larger
$\sigma$ value of the LB potential which effectively reduces the space available 
inside the IC to accommodate the H$_2$ molecule.
   
\begin{figure}
\centering
\includegraphics[width=9cm]{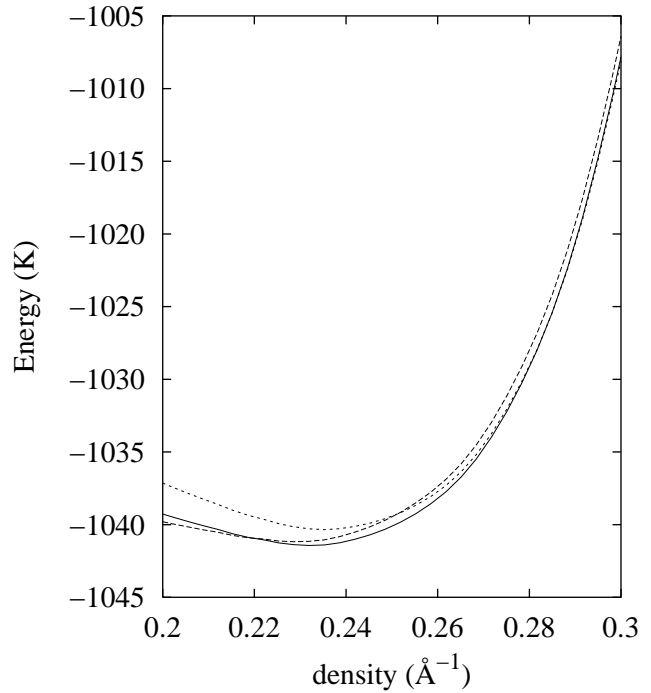}
\caption{\label{Fig 2} Same than in Fig. 1 but changing the LB C-H$_2$
potential by the WSW one. From top to bottom at $\lambda = 0.2$ 
\AA$^{-1}$, lattice dilations of 0.06, 0.05 and 0.04 \AA.
}
\end{figure}

%\begin{table}         
\begin{table*}
\centering
\begin{tabular}{ccllccc}

Radius (\AA)  & IC & $V_{\text{C-H$_2$}}$ & $V_{\text{H$_2$-H$_2$}}$ &
$(D-D_0)$ (\AA) &  $\lambda_0$ (\AA$^{-1}$) &  $e_0$ (K) \\ 
\colrule
6.9   & T &  WSW & KCLDJ  &  0.04  &  0.2087(3)  &  -1032.98(1)     \\
6.9   & T &  LB  & KCLDJ  & 0.16   &  0.2667(1)  &  -481.42(2)     \\
6.9   & T &  WSW & SG     & 0.05   &  0.2322(2)  &  -1041.41(1)    \\		
6.9   & T &  LB  & SG     &  0.17  &  0.2685(1)  &  -491.64(1)      \\
6.9$^*$ &   T &  LB & SG  &  0.11  &  0.276(12)  &  -300.35(1)      \\
6.9  & R &  WSW &  KCLDJ  &  0.05  &  0.2252(1)  &  -980.07(1)     \\
6.9  & R &  LB  &  KCLDJ  & 0.16   &  0.2662(1)  &  -422.29(2)    \\
6.9  & R &  WSW  & SG     & 0.05   &  0.2323(3)  &  -988.59(1)      \\
6.9  & R &  LB  & SG      & 0.16   &  0.2685(1)  &  -432.17(1)       \\
\colrule
6.8  &  T &  WSW  & KCLDJ &  0.01  &  0.1311(6)  &  -1096.35(1)       \\ 
6.8   & T &  LB &  KCLDJ  & 0.10   &  0.2472(1)  &  -731.55(2)        \\
6.8$\dagger$  &  T &  WSW &  SG    &  0.01  &  0.2194(5)  &  -1102.31(2)        \\
6.8  &  T  &  LB  &  SG   & 0.01   &  0.2514(2)  &  -740.64(2)        \\
6.8  &  R  &  WSW & KCLDJ & 0.01   &  0.1315(2)  &  -1045.85(1)        \\
6.8  &  R  &  LB  & KCLDJ &  0.11  &  0.2501(5)  &  -679.27(1)        \\
6.8  &  R  &  WSW & SG    &  0.02  &  0.223(1)   &  -1047.39(2)        \\
6.8  &  R  &  LB  & SG    & 0.10   &  0.2502(9)  &  -679.53(3)      \\
\end{tabular}
\caption{Dilation ($(D-D_0)$), equilibrium densities ($\lambda_0$) and
energies $e_0$ for different
radius, H$_2$-H$_2$ and C-H$_2$ interactions, and geometries (T and R). 
Figures in parenthesis are statistical errors. }
%\end{table}
\end{table*}

In order to determine if a dilation of the bundle is energetically
preferred, a series of calculations using Eq. (\ref{ener1d}) has been
carried out. As commented before, our aim is to quantitatively determine the
influence  of the potentials, the CN radius, the
elastic constant $k$, and the geometry model for the IC on a possible dilation. 
The direct output
of Eq. (\ref{ener1d}) is the energy per particle as a function of the
linear density $\lambda$ and $D$. For a given dilation $D -D_0 \geq 0$,
with fixed $D_0=17$ \AA, $e(\lambda)$ can be obtained and from it the
equilibrium point ($\lambda_0$, $e_0$) corresponding to zero pressure.
Illustrative outputs of this procedure are shown in Figs. 1 and 2.
Fig. 1 contains the H$_2$ equation of state in the IC for values of the
lattice dilation $D-D_0$ = 0.18, 0.17, 0.16, and 0.15 \AA. All the curves
have been calculated assuming a CN radius of 6.9 \AA, the SG H$_2$-H$_2$  
potential, $k = 1740$ K \AA$^{-3}$, the LB C-H$_2$  potential, and a T channel. 
The lowest energy per particle at the minimum is obtained for a dilation 
0.16 \AA, nearly coincident with the result reported in Ref. \onlinecite{Cole1}
(0.166 \AA) obtained with the H$_2$-H$_2$ KCLDJ potential and an R channel.  
If the  C-H$_2$ potential is the WSW model, the H$_2$ equation of state changes
dramatically and the dilation becomes much smaller. This is shown in Fig. 2
which differs from Fig. 1 in the C-H$_2$ potential, the other inputs
being unchanged. The curves now correspond to dilations $D-D_0$ =  0.06,
0.05, and 0.04 \AA. The minimum corresponds to a swelling of 0.05 \AA. The
equilibrium density is $\lambda_0=0.232$ \AA$^{-1}$, slightly inferior to
the one obtained with the LB potential, $\lambda_0=0.268$ \AA$^{-1}$. The
magnitude of the energy per particle is very different in the two cases:
going from LB to WSW the energy increases in a factor two. The major part
of this increment comes from the differences in the binding energies of a
single molecule $\epsilon_0$ (Table I).

The full set of results combining CN radius, C-H$_2$ and H$_2$-H$_2$
interactions, and IC geometry is reported in Table II. $\lambda_0$ 
corresponds to the H$_2$ equilibrium density at the energetically
preferred swelling, reported also in the Table. Due to the large
uncertainty in the value of $k$, we only report the influence of its
change ($3015 $ K \AA$^{-3}$ instead of  $1740$ K \AA$^{-3}$)  in the most
favorable cases for the bundle swelling.  Inspection of the data
shows that neither the H$_2$-H$_2$ potential nor the detailed form of the
IC, R or T, are relevant in the swelling process. Maintaining fixed the
radius to a value 6.9 \AA\ and the LB C-H$_2$ interaction, the dilation comes
out 0.16-0.17 \AA, in overall agreement with Calbi \textit{et al.}.
\cite{Cole1} However, if the LB interaction is substituted by the WSW one,
the swelling observed is reduced by a factor of three (from
0.17 \AA\ to 0.05 \AA), and the equilibrium density $\lambda_0$ decreases
by 15 \%.

\begin{figure}
\centering
\includegraphics[width=9cm]{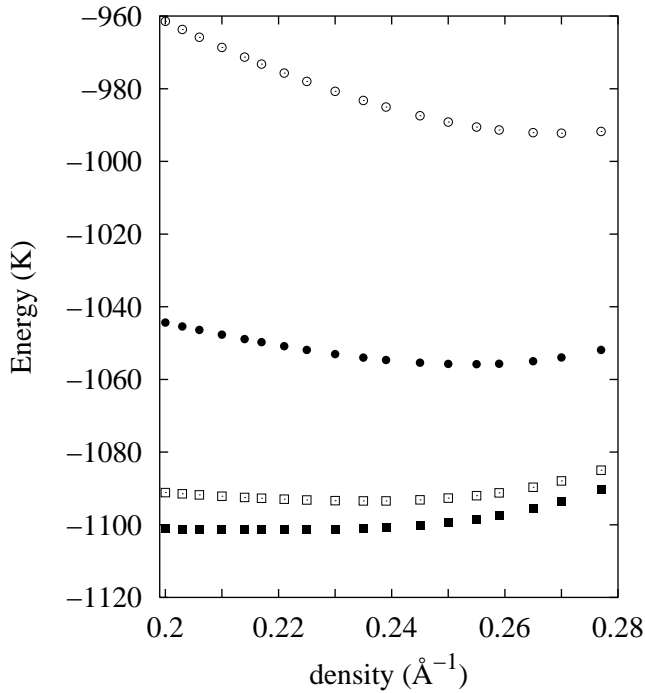}
\caption{\label{Fig 3} Full DMC calculation of molecular hydrogen 
adsorbed in the interchannels of a CN bundle. No dilation, full squares; 
$D-D_0$ = 0.05 \AA, empty squares; $D-D_0$ = 0.10 \AA, full circles; $D-D_0$ = 
0.15 \AA, empty circles. 
}
\end{figure} 
 
The second part of Table II contains a similar analysis for a CN
radius of 6.8 \AA. A general trend arising from the comparison between 
the first half of the Table (6.9 \AA)
and the second one (6.8 \AA) is the increase of the dilation with the CN
radius when the rest of the inputs are kept fixed.
This seems quite obvious since, with a smaller nanotube and the same
CN-CN distance, the room for adsorption increases. 
Thus, the swelling decreases from 0.17  to  0.11 \AA\ with the
LB C-H$_2$ interaction whereas it turns absolutely negligible in case of
using the probably most accurate WSW potential. On the other hand,
$\lambda_0$ is observed to systematically decrease when the radius is moved
from 6.9 to 6.8 \AA. Finally, the influence of the value of $k$ appears in
the Table marked with an asterisk for a particular case in which the
dilation is one of the largest values. As  it could be expected, 
$(D-D_0)$ decreases if $k$ increases: using $k= 3015 $ K \AA$^{-3}$ instead of 1740 K \AA$^{-3}$
the dilation is reduced by  30 \%. Similar reductions would be obtained in
the other cases.

\section{FULL DMC CALCULATION}

In the preceding Section it has been assumed that radial and longitudinal
degrees of freedom of H$_2$ inside IC's are not coupled. The correlations
between H$_2$ molecules were purely longitudinal and the interaction with
the surrounding walls was solved only for the one-body problem. In this
Section, we check the validity of that approximation by making an exact
three-dimensional DMC calculation of H$_2$ in the IC. The CN bundle
cohesion term and the mean field contribution  $h(\lambda,D)$ are summed up
to the DMC energy as in Eq. (\ref{ener1d}).        

\begin{figure}
\centering
\includegraphics[width=9cm]{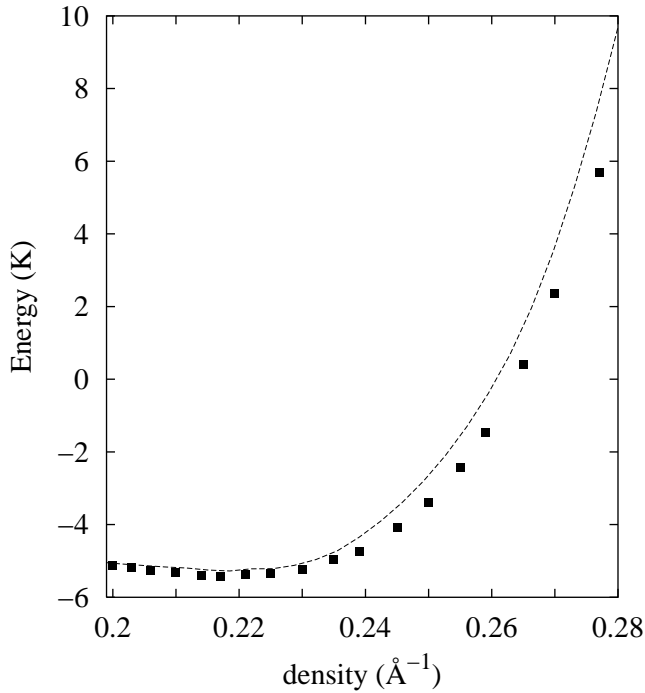}
\caption{\label{Fig 4} Comparison between a DMC calculation (full squares)
and a 1D approximation (solid line) in a case  without dilation. See
comments in the text. 
}
\end{figure}

The trial wave function for importance sampling is written as
\begin{equation}
\psi ({\bm R}) = \psi^{\text{1D}}({\bm R}) \, \psi^{\text{IC}}({\bm R}) \ ,
\label{trialfull}
\end{equation}
where $\psi^{\text{1D}}({\bm R})$ and $\psi^{\text{IC}}({\bm R})$ 
are the same than in the 1D calculation, Eqs. (\ref{trial1d}) and
(\ref{trialic}), respectively. The optimal values for the parameters
entering Eq. (\ref{trialfull}) are the same than the ones in the preceding
Section. The DMC calculation has been carried out for a selected case which
we consider contains the most reasonable model. The CN radius is 6.8 \AA\
and the IC is of T type. The H$_2$-H$_2$ interaction is SG and the
parameter set for C-H$_2$ is the WSW one.

In Fig. 3, the total energy per particle of H$_2$ in the IC is shown as a
function of $\lambda$ and for different bundle lattice values, 
$D-D_0$ = 0, 0.05, 0.10, and 0.15  \AA. The minimum of the energy is
achieved with $D=D_0$, what implies an absence of
dilation in the bundle of nanotubes. This is in agreement with the
corresponding case in Table II, marked by a dagger. In the full 3D 
calculation the equilibrium density is $\lambda_0 = 0.2184(12)$ \AA$^{-1}$, 
to be compared with $0.2194(5)$ \AA$^{-1}$ from
Table II. The second value is within the error bar of the first one.
The energy per hydrogen molecule at the equilibrium density is
also very similar in both cases, 
$e_0 = -1101.43(1)$ K and  $e^{\text{1D}}_0 = -1102.31(2)$  K. 
This clear overlap means that the 1D model (\ref{ener1d})  is a very good 
approximation to the physics of H$_2$ inside
the interchannels of a bundle of nanotubes around the equilibrium density. 

Finally, in Fig.4 the comparison between both calculations is extended to
higher densities. In the energy scale, the binding energy $\epsilon_0=
-1096.02$ K (Table I) is now subtracted. Equation (\ref{ener1d}) provides un
upper bound to the exact result which is very close to the exact energies
near $\lambda_0$. However, its quality worsens when the density increases
due to the emergence of the radial degree of freedom which is frozen in
the 1D model.

\section{SUMMARY AND CONCLUSIONS}

The equation of state of H$_2$ adsorbed in the IC's of a CN bundle has been
calculated using microscopic theory. By means of the DMC method, which
provides exact results for given model potentials, the possible dilation
of the bundle induced by adsorption has been carefully analyzed. The most
complete analysis has been done in the 1D model.\cite{Cole1} In
spite of its simplicity, this approach (\ref{ener1d}) has proven to be very accurate when 
compared with a full 3D calculation, specially near the equilibrium density. 
Playing
with the different alternatives for the parameters entering into the
calculation the influence of each one has been established. 
Summarizing, the results show that the critical one is the particular C-H$_2$ 
interaction. The  deeper and probably more realistic WSW potential
reduces the swelling predicted by the LB model in a significant way. The
same can be concluded about the C-C interaction, represented
here by the $k$ parameter, but to a lesser extent. 
On the other hand, sizeable changes in the H$_2$-H$_2$ interaction do not
modify the physics of the dilation process.
Our results do not
exclude completely the possibility of some lattice dilation but if it exists
its value seems clearly smaller than the figures quoted in Ref.
\onlinecite{Cole1}.

A key point in the discussion is the issue of what C-H$_2$
potential is more realistic. To this end, one should compare 
theoretical data with  experimental results.
However, the latter are really scarce. Recently, Vilches and collaborators 
\cite{vilches1,vilches2} reported data  on H$_2$ isotherms on bundles of 
closed-capped carbon nanotubes for several
temperatures. At very low coverages, they deduced the existence of H$_2$
adsorption on the ridges created between two nanotubes in the outer part
of the bundle and/or in the IC's. 
For these sites, they estimated a binding energy 
$\sim$ 700 K, 1.5 times larger than the one for molecular
hydrogen on graphite. 
Comparison  with Table I rules out the case with R= 6.9 \AA\
and the LB potential: its $\epsilon_0$ is too small.
On the other hand, the value for R = 6.8 \AA\  and the LB potential 
 seems to fit perfectly well the experimental result. 
However, there is a problem: the binding energy 700 K is supposed to
be an average of the energies of hydrogen adsorbed on the ridges
and in the IC's. Then, the real binding
energy in IC's should be greater than the experimental finding, 
fitting well with the results obtained with the WSW potential.

\begin{acknowledgments}
We are grateful to Ari Mizel, Oscar E. Vilches, and Milton W. Cole for
helpful discussions. This work was supported by DGI (Spain) Grant N$^0$
BFM2002-00466  and BQU2001-3615-C02-01,
and Generalitat de Catalunya Grant N$^0$ 2001SGR-00222.
\end{acknowledgments}

\end{document}